\mathchardef\mchQUOTE="7027
\newcommand\acca[1]{{\stackrel{\mchQUOTE\:}{\smash{\acute#1}}}}
\def\abacus{\acca\alpha\beta\alpha\xi}
\def\bfabacus{\boldmath{$\abacus$}}
\def\qabac{\ket{\mbox{$\acca\alpha$}}}
\def\quabacus{\ket{\abacus}}

\let\ox=\otimes
\newcommand{\hc}{{*}}
\newcommand{\mx}[6]{{\left#1\begin{array}{rr}#2&#3\\#4&#5\end{array}\right#6}}

\newcommand{\Mx}[3]{\left#1\begin{array}{rrrr}#2\end{array}\right#3}

\newcommand{\ket}[1]{| #1 \rangle}
\newcommand{\bra}[1]{\langle #1 |}

\newcommand{\brkt}[2]{\langle #1 \mid #2 \rangle}
\newcommand\0{\ket{0}}
\newcommand\1{\ket{1}}
\newcommand\tr{\mbox{\rm trace}}
\newcommand\Hil{\mathcal H}
\newcommand\C{\mathbb C}
\newcommand\Tz{\mathsf T}

\newcommand{\BIG}[3]{\left#1{#2}^{\vphantom{-}}\right#3}
\newcommand{\Op}[1]{\mathbf{#1}}
\newcommand\Sy{\mathsf S}
\newcommand\pd[1]{\frac{\partial}{\partial #1}}
\newcommand\Cl{\mathbb{C}\mathrm{l\!l}}
\newcommand\ui[1]{{{\scriptscriptstyle\!(\!}\mathnormal{#1}{\scriptscriptstyle\!)}}}
\newcommand\GE{\geqslant}
\newcommand\LE{\leqslant}

\newcommand\eq[1]{Eq.~(\ref{#1})}
\newcommand\Sec[1]{Sec.~\ref{sec:#1}}

\title{Algebra, Logic and Qubits: Quantum $\abacus$.}
\author{\em A. Yu.\ Vlasov}
\date{Jan--Feb 2000}
\documentclass[12pt]{article}
\usepackage{amssymb}
\begin{document}
\maketitle
\centerline{\tt quant-ph/0001100}
\begin{abstract}
  The canonical anticommutation relations (CAR) for fermion systems can be
represented by finite-dimensional matrix algebra, but it is impossible for
canonical commutation relations (CCR) for bosons. After description of more
simple case with representation of CAR and (bounded) quantum computational
networks via Clifford algebras in the paper are discussed CCR.  For
representation of the algebra it is not enough to use quantum networks
with fixed number of qubits and it is more convenient to consider Turing
machine with essential operation of appending new cells for description of
infinite tape in finite terms --- it has straightforward generalization for
quantum case, but for CCR it is necessary to work with symmetrized version of
the quantum Turing machine.  The system is called here {\em quantum abacus}
due to understanding analogy with the ancient counting devices ($\abacus$).
\end{abstract}

\section{Introduction}\label{sec:intro}

In his first article about quantum computers \cite{deutsch:turing}
David Deutsch wrote in relation with Feynman's work \cite{feynman:simul}:
{\small
\begin{quote}
 Feynman ($\mathnormal{1982}$) went one step closer to a true quantum computer
with his `universal quantum simulator'. This consists of a lattice of spin
systems with nearest-neighbour interactions that are freely specifiable.
Although it can surely simulate any system with a finite-dimensional
state space (I do not understand why Feynman doubts that it can simulate
fermion systems), it is not a computing machine in the sense of this article.
`Programming' the simulator consists of endowing it by {\em fiat} with the
desired dynamical laws, and then placing it in a desired initial state.
But the mechanism that allows one to select arbitrary dynamical laws is
not modelled. The dynamics of a true `computer' in my sense must be given
once and for all, and programming it must consist entirely of preparing it
in a suitable {\em state} (or mixed case).
\end{quote}
}
The cited article (together with \cite{deutsch:gates}) forms some standard
for many works about quantum information science and here is necessary
to say few words for explanation of some difference in methods and purposes
of present paper.

 First, it maybe even more interesting to understand simulation of boson
systems, because due to infinite-dimensional state space (see \Sec{acr})
it is impossible to apply directly approach mentioned by David Deutsch.
Is it possible to extend the ideas from systems with finite number of
states to {\em countable} number (discrete spectrum)? Quantization of harmonic
oscillator can be considered as simple example --- quantum model of natural
numbers, `quantum abacus ($\abacus$)'.

 Second, the cited methods of description of quantum computations are
similar with attempts to learn classical information science only by
`wiring schemes' of processors with huge number of gates. But usually it is
efficient to consider general methods of mathematical logic, algebra
{\em etc}.\ together or even before binary numbers and logical gates.

The algebraic methods currently are common for quantum theory and if we
are interesting in boson and fermion systems, we should consider canonical
commutation and anticommutation relations.

\section{(Anti)commutation relations}\label{sec:acr}

The using {\em annihilation} and {\em creation} operators $a$, $a^\hc$ for
definition of quantum gates was introduced by Feynman
\cite{feynman:simul,feynman:comp}. The operators meet anticommutation
relation $\{a,a^\hc\} \equiv a\,a^\hc + a^\hc\,a = 1$ for one qubit. For
system with $n$ qubits we may use CAR for $n$ fermions, {\em i.e}.\ %
introduce operators $a_i$, $a_i^\hc$ for each qubit with properties:
\begin{equation}
 \{a_i,a_j\}=\{a^\hc_i,a^\hc_j\}=0,\quad \{a^{}_i,a^\hc_j\} = \delta_{ij}
\label{CAR}
\end{equation}

The algebra generated by $2n$ operators \eq{CAR} is isomorphic with
algebra of all complex $2^n \times 2^n$ matrices and so can be used for
representation of any quantum gate \cite{ay:cliff}.

In basis $\0 = \Mx({1\\0})$, $\1 = \Mx({0\\1})$ the operators can be
expressed as:
\begin{eqnarray*}
 a & = & \frac{\sigma_x + i \sigma_y}{2} \ = \ \mx(0100)  \\
 a^\hc & = & \frac{\sigma_x - i \sigma_y}{2} \ = \ \mx(0010)
\end{eqnarray*}
\begin{eqnarray*}
 a_i & = &
       {\underbrace{1 \ox \cdots \ox 1}_{n-i-1}} \ox a \ox
       \underbrace{\sigma_z \ox \cdots \ox \sigma_z}_i \\
 a_i^\hc & = &
       {\underbrace{1 \ox \cdots \ox 1}_{n-i-1}} \ox a^\hc \ox
       \underbrace{\sigma_z \ox \cdots \ox \sigma_z}_i
\end{eqnarray*}

The representation of CAR also can be expressed in more invariant and clear
way with using of Clifford algebras. Let us recall it briefly \cite{ay:cliff}.

We can start with complex Clifford algebra $\Cl(n,\C)$ with $n$ generators
$e_i$ with relations $e_i e_j + e_j e_i = 2 \delta_{ij}$. For even $n = 2k$
there are two useful properties: first, there are $k$ pairs
$a_l = (e_{2l} + i e_{2l+1})/2$, $a^\hc_l = (e_{2l} - i e_{2l+1})/2$ those
satisfy \eq{CAR}, {\em i.e}.\ CAR. Second, there is recursive construction of
$\Cl(n+2,\C)~=~\Cl(n,\C) \ox \Cl(2,\C)$: if $e^{\ui n}_i$ ($i \GE 0$) are
$n$ generators of $\Cl(n,\C)$ and $e^{\ui2}_0$, $e^{\ui2}_1$ are generators of
$\Cl(2,\C)$ with $e^{\ui2}_{01} \equiv i e^{\ui2}_0 e^{\ui2}_1$, then
\mbox{$e^{\ui{n+2}}_i = e^{\ui n}_i\ox e^{\ui2}_{01}$},
\mbox{$e^{\ui{n+2}}_{n} = \mathbf 1_{\phantom0}^{\ui n}\ox e^{\ui2}_0$},
\mbox{$e^{\ui{n+2}}_{n+1} = \mathbf 1_{\phantom0}^{\ui n}\ox e^{\ui2}_1$}
are $n+2$ generators of $\Cl(n+2,\C)$. Because $\Cl(2,\C)$ is isomorphic with
algebra of Pauli matrices, the constructions correspond to equations above if
$e^{\ui2}_0 \mapsto \sigma_x$, $e^{\ui2}_1 \mapsto \sigma_y$,
$e^{\ui2}_{01} \mapsto \sigma_z$.

\medskip

Let us now consider similar possibility for Bose particles with
{\em annihilation} and {\em creation} operators $c$, $c^\hc$
with commutation relation $[c,c^\hc] \equiv c\,c^\hc - c^\hc\,c = 1$.
\begin{equation}
 [c_i,c_j]=[c^\hc_i,c^\hc_j]=0,\quad [c^{}_i,c^\hc_j] = \delta_{ij}
\label{CCR}
\end{equation}

The relation $[A,B] = 1$ cannot be true for finite-dimensional matrices
$A, B$ because $\tr[A,B] = \tr(AB) - \tr(BA) = 0 \ne \tr(1)$. On the
other hand infinite-dimensional algebra with the property can be simple
found. For example for algebra of functions on line it may be linear
operators\footnote{{\em Cf}.\ with well known representation of $ip$ and $q$
operators in quantum mechanics.}:
\begin{equation}
\begin{array}{c}
\Op D \colon \psi(x) \mapsto \psi'(x),
\quad \Op X \colon \psi(x) \mapsto x\,\psi(x); \\{}
[\Op D,\Op X]\,\psi(x) \equiv
\Op D\BIG({\Op X\BIG({\psi(x)})})-\Op X\BIG({\Op D\BIG({\psi(x)})}) = \\
=\BIG({x \psi(x)})'- x \psi'(x) = \BIG({\psi(x) + x \psi(x)'}) - x \psi'(x) =
\Op1\, \psi(x)
\end{array}
\label{CCRdif}
\end{equation}

It is possible to define commutation relations on algebra
of square-integrable functions with scalar product:
\[
\brkt{\psi}{\varphi} =
\int\limits_{-\infty}^{+\infty}{\overline{\psi(x)}\,\varphi(x)\,\mathrm dx}
\]
Then $\Op X^\hc = \Op X$, $\Op D^\hc = -\Op D$ and so
$c = (\Op{X+D})/\sqrt{2}$ and $c^\hc = (\Op{X-D})/\sqrt{2}$ satisfy commutation
relation for simplest case of one variable $[c,c^\hc] = 1$. The operators
is also related with quantization of harmonic oscillator \cite{dau:qm} ---
there is recurrence relation for eigenfunctions $\psi_n(x)$ of the 1D oscillator:
$\psi_n(x) = c^\hc \psi_{n-1}(x)/\sqrt{n}$ and so $c^\hc$ really corresponds
to `rudimentary' creation operator {\em i.e}.\ excitation of the system on
next level.

For more general case of \eq{CCR} it is possible to define $n$ pairs of
operators $c^{}_i$, $c^\hc_i$ on algebra of functions with $n$ variables
$\psi(x_1,\ldots,x_2)$:
\begin{equation}
 c_i = \frac{1}{\sqrt{2}}\Bigl(x_i + \pd{x_i}\Bigr), \quad
 c^\hc_i = \frac{1}{\sqrt{2}}\Bigl(x_i - \pd{x_i}\Bigr)
\end{equation}

\medskip

Now let us consider more formal representation of CCR used in secondary
quantization for case of Bose statistics \cite{dau:qm}. Here is used
{\em occupation numbers} representation $\ket{\mathit\Psi}=\ket{n_1,n_2,\ldots}$
with $n_i$ is number of particles in $i$-th state.
Then operators $c_i$ and $c^\hc_i$ are formally defined as:
\begin{equation}
\begin{array}{rcl}
c_i\, \ket{n_1,n_2,\ldots,n_i,\ldots} &=&
 \sqrt{n_i}\,\ket{n_1,n_2,\ldots,n_i-1,\ldots} \\
c^\hc_i\, \ket{n_1,n_2,\ldots,n_i,\ldots} &=&
 \sqrt{n_i+1}\,\ket{n_1,n_2,\ldots,n_i+1,\ldots}
\end{array}
\label{statBose}
\end{equation}
where definition of conjugated operator meets the standard condition
{\em i.e}.\ $\bra{n_i}c^\hc_i\ket{n_i-1} = \bra{n_i-1}c_i\ket{n_i}^\hc$.

In computational terminology it can be considered as quantum version of
ancient\footnote{Very ancient, because in {\em Abriss der geschichte der
mathematik} ({\em A Brief Review of the History of Mathematics}) by
von Dirk J. Struik (Berlin 1963) is mentioned that
{\em The Rhind Mathematical Papyrus} written about 3650 years ago already
used more or less directly both decimal and binary number systems.}
counting device with `unary'\footnote{I `borrowed' the term
{\em unary} from Seth Lloyd (exchange about \cite{lloyd}), to avoid
ambiguity of word `unitary' in given context.} number system ({\em vs}.\ %
binary or decimal) --- number $n$ is represented as
`$\underbrace{\prime\prime\prime\cdots\prime\prime}_n$'.
Let us call it $\abacus$ (abacus).

\section{Quantum infinite Turing machine}\label{sec:turing}

In classical theory of recursion the Turing machine is supplied with infinite
tape. Usually it is not considered as practical linitation due to argument:
let us start with finite tape and we always may add new sections
(cells) if head of Turing machine going to reach the end of the tape.

In quantum computation instead of section of the tape with two states
we have two-dimensional Hilbert space $\Hil_2$, instead of finite tape
with $n$ sections here is tensor power of $\Hil$:
\[
 \Tz\!_n(\Hil) \equiv \Hil^{{\ox}n} = \underbrace{\Hil \ox \cdots \ox \Hil}_n
\]
Then an analogue of discussed operation may be construction of space%
\footnote{It is {\em tensor algebra} of $\Hil$.} \cite{slang:alg}:
\begin{equation}
\Tz\!_{*}(\Hil) \equiv \bigoplus_{k=1}^\infty{\Tz\!_k(\Hil)} =
\Hil \oplus (\Hil\ox\Hil) \oplus (\Hil\ox\Hil\ox\Hil) \oplus \cdots
\label{TzSum}
\end{equation}
sometime it is convenient to extend summation for $k=0$
with $\Tz\!_0(\Hil) \equiv \C$.

Here subspace $\Tz\!_k(\Hil)$ in direct sum corresponds to Turing
machine with $k$-qubits tape. We considered only tape of quantum Turing
machine, because it is enough for future description of CCR. The bounded
quantum Turing machine \cite{benioff:erase} also uses other elements
and more than one tape and in addition to the property it should be mentioned that
$\Tz\!_{*}(\Hil)$ maybe more appropriate for description of semi-infinite
Turing tape with necessity of extension of the summation in \eq{TzSum} for
negative $k$ ({\em cf}.\ also $K$-theory) or using two half-tapes for each
infinite tape.

Because it is quantum system we could have superposition of two states
with different number of qubits\footnote{In physics
may exist superselection laws those prohibit some superpositions.},
and here is important that states with different number of qubits belong
to {\em orthogonal} subspaces.

It is possible to introduce Hermitian scalar product by summation of each
component, {\em i.e}.\ for $\Psi = \sum_k\psi_{\!_k}$,
$\Phi~=~\sum_k\varphi_{\!_k}$
where $\Psi,\Phi \in \Tz\!_{*}(\Hil)$ and
$\psi_{\!_k},\varphi_{\!_k}~\in~\Tz\!_k(\Hil)$:
\begin{equation}
\brkt{\Psi}{\Phi} = \sum_{k=0}^\infty{\brkt{\psi_{\!_k}}{\varphi_{\!_k}}}
\label{usScal}
\end{equation}

\section{Symmetric qubits}\label{sec:syqubit}

Let us now consider symmetric spaces $\Sy_k(\Hil)$ \cite{slang:alg,manin:lag}.
For two-dimensional $\Hil$ with basis $e_0$, $e_1$ the space is produced
by $k+1$ elements $e_{00\ldots00}$, $e_{00\ldots01}$, \ldots,
$e_{01\ldots11}$, $e_{11\ldots11}$. Let us use notation $e_{\{i,k-i\}}$ for
the elements: $e_{\{k,0\}}$, $e_{\{k-1,1\}}$, \ldots, $e_{\{0,k\}}$.
It could be enough to use only one index $e_{\{i\}}$, $i=0,\ldots,k$ if
we would work with space $\Sy_k(\Hil)$, but it is not enough for
$\Sy_{*}(\Hil)$ defined as:
\begin{equation}
\Sy_{*}(\Hil) \equiv \bigoplus_{k=1}^\infty{\Sy_k(\Hil)} =
\Hil \oplus (\Hil\odot\Hil) \oplus (\Hil\odot\Hil\odot\Hil) \oplus \cdots
\end{equation}
where `$\odot$' is used for symmetric product.

Due to isomorphism of symmetric space $\Sy_k$ with space of homogeneous
$k$-polynomials \cite{slang:alg} elements of $\Sy_k(\Hil)$ can be
represented as homogeneous polynomials with two variables $\xi$, $\eta$:
$p_k(\xi,\eta) = a_k \xi^k + a_{k-1} \xi^{k-1}\eta + \cdots + a_0 \eta^k$ and
then element $\Sy_{*}(\Hil)$ corresponds to arbitrary (non-homogeneous) polynomial
with two variables $p(\xi,\eta)$; $e_{\{i,j\}} \mapsto \xi^i\eta^j$.

It should be mentioned also, that $\Sy_k(\Hil)$ can be considered as
space of polynomials with one variable and with degree less or equal than $k$
{\em i.e}.\ \mbox{$p_k(\zeta) = a_k \zeta^k + a_{k-1} \zeta^{k-1} + \cdots + a_0$}
and there is special case with one or more higher coefficients are zeros
$a_i=0$, $k \GE i > l$. If $\zeta_1,\ldots,\zeta_l$, $l\LE k$ are roots of the
polynomial, then factorization of the $p_k(\zeta)=a_l\Pi_{i=1}^l(\zeta-\zeta_i)$
corresponds to factorization of $p_k(\xi,\eta)$ on $k$ terms:
\[
a_k \xi^k + a_{k-1} \xi^{k-1}\eta + \cdots + a_0 \eta^k =
(\alpha_1\xi - \beta_1\eta) \times \cdots \times (\alpha_k\xi - \beta_k\eta)
\]
where $\alpha_i\zeta_i=\beta_i$ and maybe $\alpha_i = 0$ if $l < k$ and
$i > l$ {\em i.e}.\ formally $\zeta_i = \infty$, but here is more rigorously
to use projective spaces\footnote{Projective spaces make possible to
get rid of special cases like $\zeta_i=\infty$ if $a_k \cdots=0$.}
--- spaces of rays in terminology more usual for quantum mechanics.
Projective coordinate $\zeta$ corresponds to ray $(\xi,\eta) \sim
(\lambda\xi,\lambda\eta)$ in $\Hil$ or point on Riemann (Bloch) sphere
and roots $\zeta_i$ correspond to pairs $(\alpha_i,\beta_i) \sim
(\lambda\alpha_i,\lambda\beta_i)$.

Due to the property element of $\Sy_k(\Hil)$ up to multiplier is defined
by $k$ points $(\alpha_i,\beta_i)$ from $\Hil$:
$\Pi_{i=1}^k{(\alpha_i\xi - \beta_i\eta)} = \lambda p_k(\xi,\eta)$
and because multiplication of each pair
$(\alpha_i,\beta_i) \mapsto (\lambda_i\alpha_i,\lambda_i\beta_i)$ changes
only common multiplier $\lambda$ for same $p_k$, it is correct map from
$k$ rays in $\Hil$ to ray in $\Sy_k(\Hil)$.

So, symmetric qubits are never entangled, each element of $\Sy_k(\Hil)$
can be represented as symmetrical product of $k$ qubits, elements of $\Hil$.
Because $\Sy_k(\Hil)$ is $k+1$-dimensional linear space and can be used as
space of states for particle with spin $k/2$, the factorization described before
explains why state of the particle always can be described as $k$ points on
Riemann sphere\footnote{There is popular introduction and two references in
\cite[\S6; Objects with large spin]{penrose:emp}.}.

\medskip
{\small
\noindent\underline{Note:} the model of symmetric qubits may looks like
some violation of Pauli's spin-statistics principle. Really, one qubit can be
considered as fermion, especially in the context of the paper with
anticommutation relation and Pauli's matrices. It should be said for
justification, that it is very convenient mathematical model of spin-$\frac{k}2$
system originated by Weyl, Majorana, Penrose and the symmetric spin-$\frac12$
subsystems can be considered as formal math `ghosts' (or `colored' like
quarks). Yet another reason --- the qubit is an abstract two-states system
and it is not quite correct to talk about spins and statistics, it maybe
electron with spin half or photon with spin one, or some model described by
Schr\"odinger equation with potential well, {\em i.e}.\ by one-component,
scalar wave function. And next, the coordinate dependence is not considered
in usual models of qubit and so Pauli's exclusion principle sometime can
be formally avoided by suggestion about different locations for each qubit.
{\em Cf}.\ also misc.\ 2 at end of \Sec{qubacus}.
}

\medskip

Now let us define scalar product on $\Sy_k$ and $\Sy_{*}$. It is convenient
together with basis $e_{\{i,j\}}$, $i+j=k$ to consider:
\begin{equation}
 \tilde{e}_{\{i,j\}} \equiv \frac{e_{\{i,j\}}}{\sqrt{i! j!}} =
 \sqrt{\frac{C^i_k}{k!}}e_{\{i,j\}}
\end{equation}

The basis is convenient by following reasons: First, in `more physical'
definition \cite{manin:lag} the $\Sy_k(\Hil)$ is {\em subspace}
($\Sy_k \subset \Tz\!_k$) of
symmetrical tensors with operation of {\em symmetrization} by summation
of $k!$ transpositions $\sigma(T)$ of indexes for given $T \in \Tz\!_k$:
$\Sy(T)=\frac{1}{k!}\sum_{\sigma} \sigma(T)$, and if to consider
$e_{\{i,k-i\}}$ as element $e_{00\ldots11}$ of $\Tz\!_k$, then
$|\Sy(e_{\{i,k-i\}})| = 1/\sqrt{C^i_k}$ (it is sum of all $C^i_k$ possible
transpositions with coefficient $1/C^i_k$) and needs for normalizing
multiplier proportional to $\sqrt{C^i_k}$ (in \cite{slang:alg} is used second
definition of symmetric space as {\em quotient space} $\Sy = \Tz/{\mathfrak S}$,
where ${\mathfrak S}$ is equivalence relation: $T \sim \sigma(T)$).

Second, $\tilde{e}_{\{i,j\}}$ form representation of $SU(2)$ group in
$SU(k+1)$ \cite{weyl:grpqm} in such a way, that if we use other basis
$U \colon (e_0,e_1) \mapsto (e'_0,e'_1)$, $U \in SU(2)$ then
$\tilde{e}'_{\{i,j\}}$ are also connected with $\tilde{e}_{\{i,j\}}$
by some {\em unitary} transformation from $SU(k+1)$.

Let us use for $\Sy_k$ basis $\tilde{e}_{\{i,j\}}$ with standard
$\brkt{\cdot}{\cdot}$ (or $\brkt{\cdot}{\cdot}/k!$, see \eq{expScal} below)
together with $e_{\{i,j\}}$ considered as transformation to other basis with
Hermitian scalar product defined by diagonal matrix $h_{ii} = C^i_k$
(or \mbox{$h_{ii} = C^i_k/k! = \frac{1}{i!(k-i)!}$}) --- the basis is convenient
for representation of $\Sy_k$ as space of polynomials.

\smallskip

The scalar product on $\Sy_{*}(\Hil)$ may be defined as in \eq{usScal}, but
it is more convenient to use also:
\begin{equation}
\brkt{\Psi^\Sy}{\Phi^\Sy}_{\exp} = \sum_{k=0}^\infty{
  \frac{1}{k!}\brkt{\psi_{\!_k}^\Sy}{\varphi_{\!_k}^\Sy}}
\label{expScal}
\end{equation}

\section{Quantum \bfabacus}\label{sec:qubacus}

It is possible to use $\Sy_{*}(\Hil)$ as an example of {\em quantum
$\abacus$} (let us denote it as $\quabacus$ or $\qabac$) introduced in
\Sec{acr}. The approach makes possible to link it with $\Tz\!_{*}(\Hil)$ and
quantum Turing machine, $\Sy(\ket{\mbox{TM tape}})\to\quabacus$

The elements $e_{\{i,j\}}$ or $\tilde{e}_{\{i,j\}}$ of basis $\Sy_{*}(\Hil)$ can be
used as basis $\ket{i,j}$ of $\qabac$ with two different kinds of states
$n_0=i$, $n_1=j$ that can be considered also as composite system of two
$\qabac$ with infinite series of states for each one:
$\Sy_{*}(\Hil) \cong \qabac_0 \ox \qabac_1$.

It is possible to introduce operators $c_i$, $c^\hc_i$; $i=0,1$ by
\eq{statBose}. Let us $\tilde{e}_{\{n_0,n_1\}} \equiv \ket{n_0,n_1}$. Then:
\begin{equation}
{\setlength{\arraycolsep}{.1em}
\begin{array}{rclrcl}
c_0\, \ket{n_0,n_1} &=&
 \sqrt{n_0}\,\ket{n_0-1,n_1}, &
\ c^\hc_0\, \ket{n_0,n_1} &=&
 \sqrt{n_0+1}\,\ket{n_0+1,n_1} \\
c_1\, \ket{n_0,n_1} &=&
 \sqrt{n_1}\,\ket{n_0,n_1-1}, &
\ c^\hc_1\, \ket{n_0,n_1} &=&
 \sqrt{n_1+1}\,\ket{n_0,n_1+1}
\end{array}
}
\label{statAbac}
\end{equation}
Here numbers of zeros and units $n_0, n_1$ are used instead of
$n_1, n_2, \ldots$ 

\smallskip

As understanding example of such system it is possible to use two-dimensional
oscillator:
\[
i \hbar\dot\psi(x,y,t) = \left(\frac{m\omega^2}{2}(x^2 + y^2) -
\frac{\hbar^2}{2m}\Delta_{x,y}\right) \psi(x,y,t)
\]

If $\phi_k(x)$, $k \GE 0$ is stationary solution of one-dimensional oscillator
for energy $E=(k+\frac{1}{2})\hbar\omega$, then for 2D oscillator function
$\phi_k(x)\phi_j(y)$ is solution for $E=(k+j+1)\hbar\omega$ and so for
any natural $n \GE 0$ there is $n+1$ dimensional space of solutions
for given energy $E=(n+1)\hbar\omega$:
\[
\phi_n(x,y) = \sum_{k=0}^n {\alpha_k\, \phi_k(x)\, \phi_{n-k}(y)}
\]
and nonstationary solution has form:
\[
\phi(x,y,t) = \sum_{n=0}^\infty {A_n \phi_n(x,y) e^{i\,(n+1) \hbar\,\omega\,t} }
\]

The example shows, how tensor product of two infinite-dimensional spaces
$\Hil_\infty\ox\Hil_\infty$ is decomposed on direct sum of linear spaces
with dimensions $1,2,3,\ldots$ for the simple case in good
agreement with formal mathematical constructions discussed above.

\medskip

It is useful to consider \eq{statAbac} in basis $e_{\{n_0,n_1\}}$:
{\samepage
\begin{equation}
{\setlength{\arraycolsep}{.1em}
\renewcommand{\arraystretch}{1.25}
\begin{array}{rcl}
c_0\, e_{\{n_0,n_1\}} &=&
{c_0\ket{n_0,n_1}}{\sqrt{n_0!n_1!}} =
{\sqrt{n_0}\,\ket{n_0-1,n_1}}{\sqrt{n_0!}\sqrt{n_1!}} \\ &=&
{n_0\,\ket{n_0-1,n_1}}{\sqrt{n_0{-}1\,!}\sqrt{n_1!}} =
n_0\, e_{\{n_0-1,n_1\}} \\
c^\hc_0\, e_{\{n_0,n_1\}} &=&
{c^\hc_0\ket{n_0,n_1}}{\sqrt{n_0!n_1!}} =
{\sqrt{n_0+1}\,\ket{n_0+1,n_1}}{\sqrt{n_0!}\sqrt{n_1!}} \\ &=&
{\ket{n_0+1,n_1}}{\sqrt{n_0{+}1\,!}\sqrt{n_1!}} =
e_{\{n_0+1,n_1\}}
\end{array}
}
\label{CCRtilde}
\end{equation}
and similarly with $c_1, c^\hc_1$ and $n_1$.}

The equations \eq{CCRtilde} demonstrate relation between CCR in secondary
quantization \eq{statBose} with CCR in differential algebra \eq{CCRdif}.
Really, let us consider polynomials with two variables $\chi_0, \chi_1$,
then $e_{\{i,j\}} \mapsto \chi_0^i \chi_1^j$,
$c_i\, p(\chi_0,\chi_1) = \pd{\chi_i} p(\chi_0,\chi_1)$,
$c^\hc_i\, p(\chi_0,\chi_1) = \chi_i \cdot p(\chi_0,\chi_1)$.

\subsection*{Miscellany}

{\bf 1.} Let us return to initial notation ($\xi=\chi_0$, $\eta=\chi_1$) for
polynomial basis $e_{\{i,j\}} \leftrightarrow \xi^i\eta^j$  defined
earlier in \Sec{syqubit}. The linear operators $c_0$ and $c_1$ are isomorphic
with two partial derivatives $\partial_\eta$ and $\partial_\xi$ and have some
interesting property, `linear merging' (`anti-cloning').

It was described in \Sec{syqubit} that state of symmetric qubits up to
multiplier, {\em i.e}.\ ray in $\Sy_n(\Hil)$, can be described by $n$ points
on Riemann sphere. The operators $\partial_\xi$ and $\partial_\eta$ can be
considered as maps $\Sy_n(\Hil) \to \Sy_{n-1}(\Hil)$ and also between spaces
of rays --- from sphere with $n$ marked points $(\zeta_1,\ldots,\zeta_n)$ and
without one pole (the pole maps to zero) to sphere without same pole and with
$n-1$ marked points $(\zeta'_1,\ldots, \zeta'_{n-1})$. For $n=2$ they map
$\Sy_2(\Hil) \to \Hil$ and $(\zeta_1,\zeta_2) \mapsto (\zeta'_1)$. If two
points on Riemann sphere coincide $\zeta_1=\zeta_2$, then due to standard
property of differential we have $\zeta'_1 = \zeta_1 = \zeta_2$ and so we have
maps $(\zeta_1,\zeta_1) \mapsto (\zeta_1)$ described by {\em linear}
operators $\partial_\xi$ or $\partial_\eta$.

Cloning is suggested may not to be linear, but for symmetric qubits `opposite'
operation may be defined as linear one almost everywhere except of one point
and it may be arbitrary point of Riemann sphere because we can use operator
$\vec\partial_v \equiv v_1 \partial_\xi + v_2 \partial_\eta$.

\medskip

\noindent{\bf 2.} Let us consider even subspace of $\Sy_{*}(\Hil)$ defined as:
$\Sy^2_{*}(\Hil) \equiv \bigoplus_{k=1}^\infty{\Sy_{2k}(\Hil)}$ and
$\Tz^2_{*}(\Hil) \equiv \bigoplus_{k=1}^\infty{\Tz\!_{2k}(\Hil)}$.
Such spaces may be more appropriate for taking into account the
Pauli's exclusion principle, but here is discussed only simplest
illustrative example.

Let us introduce operators $\tilde{\mathbf D} \equiv c_0\,c_1$ and
$\tilde{\mathbf X} \equiv \xi\,\eta$
{\em i.e}.\ :
\[
\tilde{\mathbf D}\, \ket{n_0,n_1} = \sqrt{n_0 n_1}\,\ket{n_0-1,n_1-1},\quad
\tilde{\mathbf X}\, \ket{n_0,n_1} = \ket{n_0+1,n_1+1}
\]
The operators act on $\Sy^2_{*}(\Hil)$ and there is isomorphism of subspace
with basis $\ket{n,n}$, {\em i.e}.\ $n_0=n_1$ and
space of functions with operators \eq{CCRdif}, $\ket{n,n} \mapsto x^n$.

The operators are described here not only because of trivial identity
$\sqrt{n_0 n_1}~=~n$ for $n_0=n_1=n$, but as an illustrative introduction to
more difficult 4D case where momentum operator $p_i$ and differentials
$\pd{x_i}$ have {\em two} indexes as spinoral objects. But it is already away
from theme of the paper\footnote{It may be suggested, that in
\cite{feynman:simul} the words about simulation of bosons by spin lattice
are also related with same area of research {\em i.e}.\ `Feynman checkerboard
model', `Penrose spin network', `Finkelstein-Selesnick quantum net'
{\em etc}.\ \cite{check}.}.

\section{Conclusions and discussion}

In the paper is discussed algebraic approach to quantum computational models
with unlimited number of discrete states. Similarly with classical recursion
theory instead of `actual infinity' here is considered quantum analogue of
Turing machine as sequence of systems with increasing number of states.

Here is also used simplest mathematical model with linear spaces,
isomorphic to space of polynomials. It maybe defined also by canonical
commutation relations, CCR. It is still far from more rigorous models like
relativistic quantum theories of interacting fields, but it is some step in
the direction.

Usual quantum networks correspond to physical approach with S-matrix. It
is analogue of ideas \cite{deutsch:turing} cited in introduction. There are
initial state, scattering process described by `quantum black box' and final
state. In such picture we have only two `points' ({\em in $\mapsto$ out})
instead of 4D spacetime.

Because CCR are also related with algebra of smooth functions on spacetime
(see \eq{CCRdif}) the models discussed in the paper are useful possibility
to take into account some properties of temporo-spatial systems in
quantum networks approach. It is only necessary to accept
infinite or dynamically changing number of qubits.

Does the mathematical models like $\Tz\!_{*}(\Hil_2)$ with infinite sequence of
linear spaces with increasing dimensions devote an attention instead of
`actual' infinite-dimensional space $\Hil_\infty$? The paper is an attempt
to make positive answer. For example operators of
creation and annihilation can be simply expressed via transition
between spaces $\Sy_k(\Hil)$ of such sequence.

A nontrivial property of such sequences is orthogonality of any states in
different terms.  For example a state of tape  of quantum Turing
machine is orthogonal with state that differs only on extra one empty section,
but in classical case they are considered as the same. The quantum $\abacus$,
$\Sy_{*}(\Hil)$ has same property, but it is more clear from physical model
of secondary quantization, because $\ket{n_0,n_1}$ and $\ket{n_0+1,n_1}$
are obviously orthogonal. But $\Sy_{*}(\Hil)$ is simply symmetrization
$\Sy(\Tz\!_{*}(\Hil))$ and so model of quantum Turing machine as infinite
sequence of orthogonal linear spaces is not much more unusual than
2D harmonic oscillator discussed as a model of $\Sy_{*}(\Hil)$ in \Sec{qubacus}.

\section*{Acknowledgements}
I am grateful to David Deutsch for some useful exchange and for
inspiration my interest to quantum information science in relation
with reading \cite{deutsch:qtm} few years ago. Many thanks to Seth Lloyd for
interesting communication. Certainly, understanding of the particular
area of quantum mechanics would be much less effective for me without big
help of David Finkelstein with explanation and discussion about some general
principles of philosophy of the quantum World.

{

}
\end{document}